\documentclass[runningheads,a4paper]{llncs}

\usepackage[T1]{fontenc}
\usepackage[utf8]{inputenc}
\usepackage{lmodern}

\usepackage{amssymb}
\usepackage{amsmath}
\usepackage{graphicx}
\usepackage{multirow}
\usepackage{color}
\usepackage{paralist}
\usepackage{subfigure}
\usepackage{listings} 
\usepackage{multirow}
\usepackage{algorithm}
\usepackage[noend]{algpseudocode}

\usepackage[textsize=tiny]{todonotes}

\usepackage{pifont}
\usepackage{xcolor}

\makeatletter
\def\BState{\State\hskip-\ALG@thistlm}
\makeatother

\usepackage{url}
\newcommand{\keywords}[1]{\par\addvspace\baselineskip
\noindent\keywordname\enspace\ignorespaces#1}

\usepackage{listings}
\usepackage{color} 
\definecolor{grey}{rgb}{0.9,0.9,0.9} 
\begin{document}

\mainmatter 

\titlerunning{Quality Assessment of  Linked Datasets using Probabilistic Approximation}
\title{Quality Assessment of Linked Datasets using Probabilistic Approximation%
\thanks{This work is supported by the European Commission under the Seventh Framework Program FP7 grant 601043 (\texttt{http://diachron-fp7.eu}).}}

\author{Jeremy Debattista, Santiago Londo\~{n}o, Christoph Lange, S\"{o}ren Auer}
\institute{University of Bonn \& Fraunhofer IAIS \\
\email{\{debattis,londono,langec,auer\}@cs.uni-bonn.de}
}

\maketitle

\begin{abstract}
With the increasing application of Linked Open Data, assessing the quality of datasets by computing quality metrics becomes an issue of crucial importance.
For large and evolving datasets, an exact, deterministic computation of the quality metrics is too time consuming or expensive.
We employ probabilistic techniques such as Reservoir Sampling, Bloom Filters and Clustering Coefficient estimation for implementing a broad set of data quality metrics in an approximate but sufficiently accurate way.
Our implementation is integrated in the comprehensive data quality assessment framework Luzzu.
We evaluated its performance and accuracy on Linked Open Datasets of broad relevance.

\keywords{data quality, linked data, probabilistic approximation}

\end{abstract}


\section{Introduction}
\label{sec:introduction}



The Web of Data is continuously changing with large volumes of data from different sources being added.
Inevitably, this causes the data to suffer from inconsistency, both at a semantic level (contradictions) and at a pragmatic level (ambiguity, inaccuracies), thus creating a lot of noise around the data.
It also raises the question of how \emph{authoritative} and \emph{reputable} the data sources are.
Taking \emph{DBpedia}\footnote{\url{http://www.dbpedia.org}} as an example, data is extracted from a semi-structured source created in a crowdsourcing effort (i.e. Wikipedia).
This extracted data might have quality problems because it is either mapped incorrectly or the information itself is incorrect.
\emph{Data consumers} increasingly rely on the Web of Data to accomplish tasks such as performing analytics or building applications that answer end user questions.
Information overload is a consistent problem that these consumers face daily.
Ensuring quality of the data on the Web is of paramount importance for data consumers, since it is infeasible to filter this infobesity manually.

A particular challenging area is the quality analysis of large-scale, evolving Linked Data datasets.
In their editorial~\cite{journals/semweb/HitzlerJ13}, Hitzler and Janowicz claim that Linked Data is an ideal pilot to experiment with the 4\textsuperscript{th} paradigm of big data (Veracity).
However, Linked Data is frequently overlooked due to its reputation of being of poor quality.
The quality of data can usually not be described using a single measure, but commonly requires a large variety of quality measures to be computed.
Doing this for large datasets poses a substantial data processing challenge.
However, for large datasets meticulously exact quality measures are usually not required.
Instead users want to obtain an \emph{approximate} indication of the quality they can expect.

Previous work on Linked Data quality analysis primarily employed deterministic algorithms (cf. the survey by Zaveri et al.~\cite{Zaveri2012:LODQ}).
Although such algorithms usually have polynomial complexity they are intractable for large datasets and it is difficult to reach runtimes sufficient for practical applications.
The rationale of this paper is to show that we can apply probabilistic techniques to assess Linked Data quality.
In particular, we employ three techniques commonly used in big data applications: \emph{Reservoir Sampling}, \emph{Bloom Filters} and \emph{Clustering Coefficient estimation}.
We develop strategies how these techniques can be applied to boost quality metric computations.
We also thoroughly evaluate the quality metrics to tweak the required parameters for more accurate results yet keeping the running time acceptable.
All implemented quality metrics are part of a large quality assessment framework, \emph{Luzzu}\footnote{Luzzu is open source and available to download from \url{http://eis-bonn.github.io/Luzzu}.}~\cite{DBLP:journals/corr/DebattistaLLA14}.

The rest of this paper is organised as follows. 
Section~\ref{sec:relatedwork} looks at the state of the art.
Section~\ref{sec:preliminaries} provides preliminaries. 
Section~\ref{sec:datametrics} details the Linked Data quality metrics under discussion. 
Section~\ref{sec:implementation} discusses the implementation of the big data techniques and metrics.
Section~\ref{sec:experiments} reports our evaluation results.
Final remarks and conclusions are presented in Section~\ref{sec:conclusion}.




\section{State of the Art}
\label{sec:relatedwork} 

In a recent article, Dan O'Brien~\cite{obrien:online2015} discusses how big data, which is now being applied in many companies and applications, challenges data governance, including data quality.
This section overviews the state of the art in relation to the probabilistic approximation techniques that can be applied to assess data quality in Linked Open Datasets.
To our knowledge, there is currently no concrete use of such techniques to assess linked dataset quality.

Since their inception, \emph{Bloom Filters} have been used in different scenarios, including dictionaries and spell-checkers, databases (for faster join operations and keeping track of changes), caching, and other network related scenarios~\cite{DBLP:journals/im/BroderM03}.
Recently, this technique was also used to tackle the detection of duplicate data in streams in a variety of scenarios~\cite{journals/corr/abs-1212-3964,1142477,conf/fast/JainDT05,1060753}.
Such applications included the detection of duplicate clicks on pay-per-click adverts, fraud detection, URI crawling, and identification of distinct users on platforms. 
Metwally et al.~\cite{1060753} designed a Bloom Filter that applies the ``window'' principle: sliding windows (finding duplicates related to the last observed part of the stream), landmark windows (maintaining specific parts of the stream for de-duplication), and jumping windows (a trade-off between the latter two window types).
Deng and Rafiei~\cite{1142477} go a step further than~\cite{1060753} and propose the Stable Bloom Filter, guaranteeing good and constant performance of filters over large streams, independent of the streams' size.
Bera et al.~\cite{journals/corr/abs-1212-3964} present a novel algorithm modifying Bloom Filters using \emph{reservoir sampling} techniques, claiming that their approach not only provides a lower \emph{false negative rate} but is also more stable than the method suggested in~\cite{1142477}.

\emph{Random sampling}, in different forms, is often used as an alternative to complex algorithms to provide a quick yet good approximation of results~\cite{journals/toms/Vitter85}.
Sample-based approaches such as the latter were used to assess the quality of Geographic Information System data~\cite{isprs:xie2008,DBLP:dblp_journals/corr/SaberiG14}.
Xie et al.~\cite{isprs:xie2008} describe different sampling methods for assessing geographical data.
In their approach, Saberi and Ghadiri~\cite{DBLP:dblp_journals/corr/SaberiG14} sampled the original base geographical data periodically.
The authors in~\cite{DBLP:conf/icis/KaiserKH07} propose how data quality metrics can be designed to enable (1) the assessment of data quality and (2) analyse the economic consequences after executing data quality metrics.
They suggest sampling the dataset attributes to get an estimate measure for the quality of the real-world data.

Lately, various efforts have been made to estimate values within big networks, such as \emph{estimating the clustering coefficient}~\cite{conf/www/HardimanK13} or calculating the average degree of a network~\cite{Dasgupta:2014:EAD:2566486.2568019}.
Hardiman et al.~\cite{conf/www/HardimanK13} provide an estimator to measure the network size and two clustering coefficient estimators: the network average (local) clustering coefficient and the global clustering coefficient.
These measures were applied on public datasets such as DBLP, LiveJournal, Flickr and Orkut.
Similarly, Dasgupta et al.~\cite{Dasgupta:2014:EAD:2566486.2568019} calculate the average degree of a network using similar public domain datasets.
As Gu\`{e}ret et al. pointed out in~\cite{DBLP:conf/esws/GueretGSL12}, network measures can be exploited to assess Linked Data with regard to quality, as Linked Data uses the graph-based RDF data model.



\section{Preliminaries}
\label{sec:preliminaries}

The LOD Cloud\footnote{\url{http://lod-cloud.net}} comprises datasets having less than 10K triples, and others having more than 1 billion triples.
Deterministically computing quality metrics on these datasets might take from some seconds to days.
This section introduces three probabilistic techniques commonly used in big data applications; they combine with a high probability near-to-accurate results with a low running time.

\subsubsection{Reservoir sampling.}
Reservoir sampling is a statistics-based technique that facilitates the sampling of evenly distributed items.
The sampling process randomly selects $k$ elements ($\leq n$) from a source list, possibly of an unknown size $n$, such that each element in the source list has a $k/n$ probability of being chosen~\cite{journals/toms/Vitter85}.
The reservoir sampling technique is part of the \emph{randomised algorithms} family.
Randomised algorithms offer simple and fast solutions for time-consuming counterparts by implementing a degree of randomness.
Vitter~\cite{journals/toms/Vitter85} introduces an algorithm for selecting a random sample of $k$ elements from a bigger list of $n$ elements, in one pass.
The author discusses that by using a \emph{rejection-acceptance technique} the running time for the sampling algorithm improves.
The main parameter that affects the tradeoff between fast computation and an accurate result is the reservoir size ($k$).
The sample should be \emph{large enough} such that the law of large numbers\footnote{\url{http://mathworld.wolfram.com/LawofLargeNumbers.html}} can be applied.


\subsubsection{Bloom Filters.}
A Bloom Filter~\cite{journals/cacm/Bloom70} is a fast and space efficient bit vector data structure commonly used to query for elements in a set (``is element $A$ in the set?'').
The size of the bit vector plays an important role with regard to the precision of the result.
A set of hash functions is used to map each item added to be compared, to a corresponding set of bits in the array filter.
The main drawback of a Bloom Filter is that they can produce \emph{false positives}, therefore being possible to identify an item as existing in the filter when it is not, but this happens with a very low probability.
The trade-off of having a fast computation yet a very close estimate of the result depends on the size of the bit vector.
With some modifications, Bloom Filters are useful for detecting duplicates in data streams~\cite{journals/corr/abs-1212-3964}.

\subsubsection{Clustering Coefficient Estimation.}
The clustering coefficient algorithm measures the neighbourhood's density of a node.
The clustering coefficient is measured by dividing the number of edges of a node and the number of possible connections the neighbouring nodes can have.
The time complexity for this algorithm is $\mathcal{O}(n^3)$, where $n$ is the number of nodes in the network.
Hardiman and Katzir~\cite{conf/www/HardimanK13} present an algorithm that estimates the clustering coefficient of a node in a network using \emph{random walks}.
A \emph{random walk} is a process where some object jumps from one connected node to another with some probability of ending in a particular node.
A random walker stops when the \emph{mixing time} is reached.
In a Markov model, mixing time refers to the time until the chain is close to its steady state distribution, i.e. the total number of steps the random walker should take until it retires.
Given the right \emph{mixing time}, the value is proved to be a close approximate of the actual value. 
The authors' suggested measure computes in $\mathcal{O}(r) + \mathcal{O}(rd_{\max})$ time, where $r$ is the total number of steps in the random walk and $d_{\max}$ is the node with the highest degree\footnote{The number of in-links plus out-links of a node}.




\section{Linked Data Metrics}
\label{sec:datametrics}

Zaveri et al. present a comprehensive survey~\cite{Zaveri2012:LODQ} of quality metrics for linked open datasets.
Most of the quality metrics discussed are \emph{deterministic} and computable within \emph{polynomial time}.
On the other hand, once these metrics are exposed to large datasets, the metrics' upper bound grows and as a result, the computational time becomes intractable.
In this section we discuss some metrics that are known to suffer from the big data phenomenon.

\subsubsection{Dereferenceability.}
HTTP URIs should be dereferenceable, i.e. HTTP clients should be able to retrieve the resources identified by the URI.
A typical web URI resource would return a \texttt{200 OK} code indicating that a request is successful and a \texttt{4xx} or \texttt{5xx} code if the request is unsuccessful. 
In Linked Data, a successful request should return an RDF document containing triples that describe the requested resource.
Resources should either be \emph{hash} URIs or respond with a \texttt{303 Redirect} code~\cite{W3C:CoolURIs}.
The dereferenceability metric assesses a dataset by counting the number of valid dereferenceable URIs (according to these LOD principles) divided by the total number of URIs.
Yang et. al~\cite{Yang2011} describe a mechanism\footnote{Also used in the Semantic Web URI Validator Hyperthing (\url{http://www.hyperthing.org})} to identify the dereferenceability process of a Linked Data resource.

A na\"{\i}ve approach for this metric is to dereference all URI resources appearing in the subject and the object of all triples.
In this metric we assume that all predicates are dereferenceable.
This means that the metric performs at worst $2n$ HTTP requests, where $n$ is the number of triples.
It is not possible to perform such a large number of HTTP requests in an acceptable time. 


\subsubsection{Existence of Links to External Data Providers.}
This metric measures the degree to which a resource is linked to external data providers.
Ideally, datasets have a high degree of linkage with external data providers, since interlinking is one of the main principles of Linked Data~\cite{Hogan2012}.

The simplest approach for this metric is to compare the subject's resource pay-level domain (PLD) against the object's resource PLD\footnote{``PLDs allow us to identify a realm, where a single user or organization is likely to be in control.''~\cite{Muehleisen:VocabularyUsage}.
For example the PLD for \url{http://dbpedia.org/resource/Malta} is \url{dbpedia.org}.}.
Although this metric is not considered to be computationally expensive ($\mathcal{O}(n)$, where $n$ represents the number of triples), it is also a good candidate for an estimation.

\subsubsection{Extensional Conciseness.}
At the data level, a linked dataset is concise if there are no redundant instances~\cite{Mendes2012}.
This metric measures the number of unique instances found in the dataset.
The uniqueness of instances is determined from their properties and values.
An instance is unique if no other instance (in the same dataset) exists with the same set of properties and corresponding values.

The most straightforward approach is to compare each resource with every other resource in the dataset to check for uniqueness.
This gives us a time complexity of  $\mathcal{O}(i^2t)$, where $i$ is the number of instances in the datasets and $t$ is the number of triples.
The major challenge for this algorithm is the number of triples in a dataset, since each triple (predicate and object) is compared with every other triple streamed from the dataset.


\subsubsection{Clustering Coefficient of a Network.}
The clustering coefficient metric is proposed as part of a set of network measures to assess the quality of data mappings in linked datasets~\cite{DBLP:conf/esws/GueretGSL12}.
This metric aims at identifying how well resources are connected, by measuring the density of the resource neighbourhood.
A network has a high clustering cohesion when a node has a large number of neighbouring nodes, all of which are connected to each other.
This means that links may end up being meaningless~\cite{DBLP:conf/esws/GueretGSL12}.

When assessing the clustering coefficient of a network, a graph is built where the \emph{subject} and \emph{object} of a triple (either URI resources or blank nodes) are represented as vertices in the graph, whilst the \emph{predicate} is the edge between them.
As this ignores triples with literal objects, there is no direct correlation between the number of triples in a dataset and number of vertices.
Calculating this measure on a network takes $\mathcal{O}(n^3)$, especially for large datasets.
This is because each vertex in the network has to be considered:
for each vertex $v$ in the graph, we identify the number of links between the neighbours of $v$ (i.e. how many of $v$'s neighbours are connected together) and divide it by the number of possible links.

\section{Implementation}
\label{sec:implementation}

Based on the probabilistic techniques described in Section~\ref{sec:preliminaries}, we analyse how they can help in assessing quality in linked datasets.
These metrics are implemented as an extensible package for \emph{Luzzu}.
Luzzu~\cite{DBLP:journals/corr/DebattistaLLA14} is a Linked Data quality assessment framework that provides an integrated platform which: 
(1) assesses Linked Data quality using a library of generic and user-provided domain specific quality metrics in a scalable manner; 
(2) adds queryable quality metadata to the assessed datasets; and 
(3) assembles detailed quality reports on assessed datasets. 
Datasets are assessed using a sequential streaming approach.
Table~\ref{tbl:tech} shows which approximation can be used for each respective metric.

\begin{table}[tb]
\centering
\begin{tabular}{|l|l|}
\hline
\textit{\textbf{Probabilistic Approximation Technique}}                  & \textbf{Linked Data Metric}                     \\ \hline
\multirow{2}{*}{\textit{Reservoir Sampling}} & Dereferenceability                  \\ \cline{2-2} 
                                             & Links to External Data Providers    \\ \hline
\textit{Bloom Filters}     & Extensional Conciseness             \\ \hline 
\textit{Clustering Coefficient Estimation}   & Clustering Coefficient of a Network \\ \hline
\end{tabular}
\caption{Mapping Probabilistic Approximation Techniques with Linked Data Quality Metrics}
\label{tbl:tech}
\end{table}

\subsection{Reservoir Sampling}
\label{sec:impl_res_sampling}
Our implementation is based on the \emph{rejection-acceptance} technique~\cite{journals/toms/Vitter85}.
The trade-off parameter is the definition of the maximum number of items ($k$) that can be stored.
Various factors are taken to define $k$, such as the rough estimation of the size of the dataset and available memory, since this reservoir is stored in-memory.

When attempting to add an $\mathit{item}$ to the reservoir sampler, an item counter ($n$) is incremented.
This increment is required to calculate the \emph{replacement probability}, since the exact size of the source (in our case the dataset) is unknown.
The $\mathit{item}$ can be (i) \emph{added} to the reservoir, (ii) become a \emph{candidate} to replace another item, or (iii) be \emph{discarded}.
The first possible operation is straightforward. 
If the reservoir sampler has free locations ($n < k$), the $\mathit{item}$ is added.
On the other hand, when the reservoir is full, the $\mathit{item}$ can either replace another item in the list, or rejected.
The decision is made by generating a random number ($p$) between 0 and $n$.
If $p$ lies in the range of the reservoir list length (i.e. $p < k$), then the new $\mathit{item}$ replaces the current item stored in that position of the reservoir, else it is rejected.
This simulates the $k/n$ \emph{replacement probability} for all items.

\subsubsection{Estimated Dereferenceability Metric.}
Each resource URI is split into two parts: (1) the pay-level Domain (PLD), and (2) the path to the resource.
For this metric we employ a ``global'' reservoir sampler for the PLDs.
Furthermore, for each PLD we employ another reservoir sampler holding an evenly distributed sample list of resources to be dereferenced.
If the pay-level domain returns a \texttt{4xx/5xx} code upon an HTTP request, then all other sampled resources in that reservoir are automatically deemed as non-dereferenceable.
Envisaging the possibility of multiple HTTP requests to same domain or resource, we make use of the Luzzu's caching mechanism, to store HTTP requests.
The metric value is calculated as a ratio of the total number of dereferenced URIs against the total number of sampled URIs.

\subsubsection{Estimated Links to External Data Providers Metric.}
In order to measure the use of external data providers, the metric must first identify the base URI of the dataset that is being assessed.
As each triple is streamed to the metric processor, a heuristic mechanism identifies the base URI.
For this, we apply one of the two heuristics, listed in order of priority:
\begin{enumerate}
\item Extract the base URI from a triple having the \emph{predicate} \texttt{rdf:type} and \emph{object} \texttt{void:Dataset} or \texttt{owl:Ontology}.
\item The URI (PLD) with the maximum number of occurrences in the \emph{subject} of the assessed dataset.
\end{enumerate}
Each triple's \emph{object} in the dataset is then used to estimate the value of this metric, by first extracting its PLD and attempting to add it to the metric's reservoir.
The value of this metric is defined as the ratio of the number of PLDs in the sampler that are not the same as the base URI, against the total number of URIs in the sampler.

\subsection{Bloom Filters}
Linked datasets might suffer from instance duplication.
Bera et al.~\cite{journals/corr/abs-1212-3964} introduced some modifications to the mechanics of Bloom Filters to enable the detection of duplicate elements in data streams.
These modifications allow items to be inserted indefinitely by probabilistically resetting bits in the filter arrays when they are close to getting overloaded. 
The Randomised Load Balanced Biased Sampling based Bloom Filter (RLBSBF) is used to implement the detection of duplicate instances.
The authors show that this approach is efficient and generates a low \emph{false positive} rate.

An RLBSBF algorithm is initialised with (1) the total memory used by filter arrays in bits ($M$); and (2) a threshold value ($t_{\mathit{FPR}}$) for the false positive rate.
The bit vector is initialised with $k$ Bloom Filters. 
Each bloom filter has a size of $M/k$ and a hash function is mapped to it.
The authors in \cite{journals/corr/abs-1212-3964} suggest that $k$ is calculated using the threshold value $t_{\mathit{FPR}}$.
A high threshold value means faster computation but less accurate results.

\begin{figure}[tb]
  \centering
   \includegraphics[width=\textwidth]{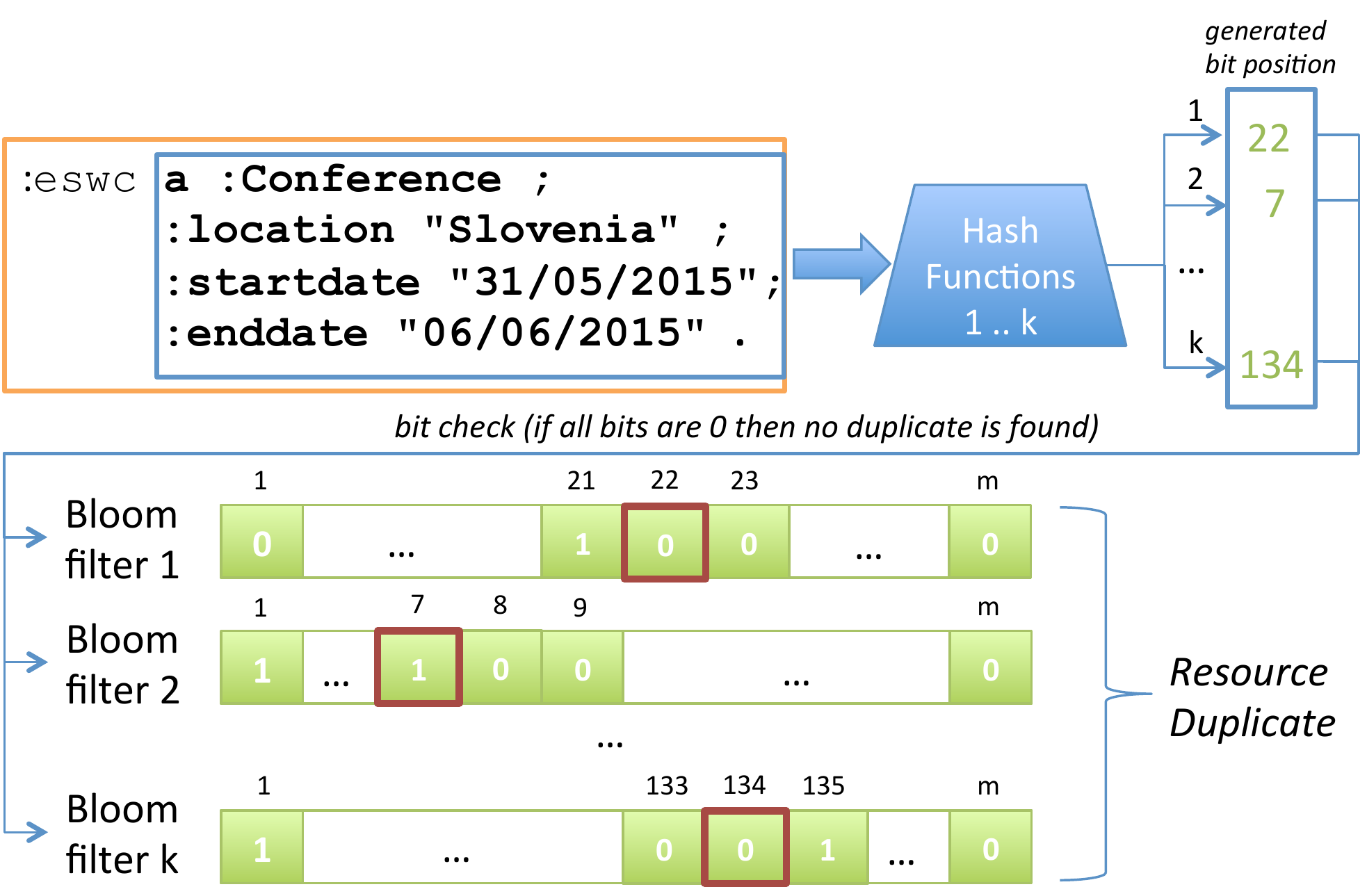} 
   \caption{Illustrating Bloom Filters with an example}
   \label{fig:bloomFilter}
\end{figure}

Whenever a new element is processed, the Bloom Filter sets all $k$ bit positions using the hash functions mapped to them.
If the bit positions were previously set in the bit vector, it means that a duplicate was detected.
Otherwise, the probabilistic resetting of bits is performed before the new element is added to the bit vector.
Our implementation uses 128-bit Murmur3\footnote{\url{https://code.google.com/p/smhasher/wiki/MurmurHash3}} hashing functions.
Figure~\ref{fig:bloomFilter} illustrates how Bloom Filters help to identify a Linked Data resource that already exists in a dataset.

\subsubsection{Estimated Extensional Conciseness Metric.}
When triples are streamed to the metric processor, the \emph{predicate} and \emph{object} are extracted and serialised as a string.
The latter string is stored in a sorted set.
This process is repeated until a triple with a different \emph{subject} identifier is processed.
The sorted set is then flattened to a string and added to the Bloom Filter, discovering any possible duplicates.
The set is then initialised again for the new resource identifier and the process is repeated until no more triples are streamed.

The main drawback of our proposed algorithm is that a dataset must be sorted by \emph{subject}, such that all triples pertaining to the same instance are streamed one after another.
Although it is common practice to publish datasets sorted by subject (e.g. DBpedia), this cannot be guaranteed in the general case.
In our experiments we pre-process RDF dumps by converting them to the N-Triples serialisation, which can be sorted by subject in a straightforward way.

\subsection{Clustering Coefficient Estimation}
In \cite{conf/www/HardimanK13}, the authors propose an approach for estimating a social network's clustering coefficient by creating a random walk.
In their proposed algorithm, Hardiman and Katzir use $\log^2n$~\footnote{the square of $\log n$} as the base \emph{mixing time}, i.e. the number of steps a random walker takes until it converges to a steady-state distribution.
However, different network characteristics lead to different \emph{mixing time}s, where well-connected networks have a small (fast) mixing time~\cite{Mohaisen2010}.

To calculate an estimate of the clustering coefficient given a random walk $R = \{x_1, x_2, \dots , x_r\}$, Hardiman and Katzir propose the Estimator~\ref{eq:clustcoeff}: 

\newtheorem{Def1}{Estimator}
\begin{Def1}
\begin{align*}
 \Phi_l = \frac{1}{r-2} \sum_{r-1}^{k=2}\phi_k \frac{1}{d_{x_k} - 1} \\
 \Psi_l = \frac{1}{r} \sum_r^{k=1} \frac{1}{d_{x_k}} \\
 \hat{c}_l \triangleq \frac{\Phi_l}{\Psi_l}
 \end{align*}
\label{eq:clustcoeff}
\end{Def1}
where $r$ is the total number of steps in the random walk $R$, $x_k$ is the index of the $k^{th}$ node in the random walk, $d_{x_k}$ is the degree of node ${x_k}$ and $\phi_k$ represents the value in the adjacency matrix $A$ in position $A_{x_{k-1},x_{k+1}}$.

\subsubsection{Estimated Clustering Coefficient Metric.}
When triples are streamed into the metric, the vertices are created by extracting the \emph{subject} and the \emph{object}, whilst the \emph{predicate} acts as a directed edge between the two nodes.
We use URI resources and blank nodes to create the network vertices.
To calculate the estimated clustering coefficient value, a random walk is performed on the graph.
Similarly to the approach in \cite{conf/www/HardimanK13}, we view the graph as undirected.
The idea is that if the random walker ends up in a dead-end (i.e. cannot move forward), it can go back to continue crawling the network.
Our mixing time parameter is $m\log^2n$.
Since linked open data advocates interlinking and re-use of resources, we expect\footnote{We are currently performing research on the mixing time of the linked datasets available in the LOD Cloud.} that such datasets have a low mixing time. 
The multiplier factor $m$ thus enables us to increase or decrease the mixing time as required.
The reason behind this is to enable a parameter modifier to the base mixing time ($\log^2n$), since it is difficult to find a one size fits all mixing time.
Estimator~\ref{eq:clustcoeff} is used to obtain a close estimate of the dataset's clustering coefficient.
Finally, the estimated value is normalised as described in~\cite{DBLP:conf/esws/GueretGSL12}.




\section{Metric Analysis and Experiments}
\label{sec:experiments}

Having implemented the metrics using probabilistic approximation techniques, we measure the computed quality metric values and runtime for the approximate metrics and compare them with the actual metrics.
For each approximate metric, we experimented with different parameter settings to identify the best parameter values.
All tests are run on a Unix virtual machine with an Intel Xeon 3.00\,GHz, with 3 cores and a total memory of 3.8\,GB.
We used a number of datasets of varying sizes and covering different application domains.
We found them on Datahub, looking for datasets tagged with the \emph{lod} tag.
These are:
\begin{itemize}
\item Learning Analytics and Knowledge (LAK) Dataset $\approx$ 75K triples;
\item Lower Layer Super Output Areas (LSOA) $\approx$ 280K triples;
\item Southampton ECS E-Prints Dataset $\approx$ 1M triples;
\item WordNet 2.0 (W3C) Dataset $\approx$ 2M triples;
\item Sweto DBLP Dataset $\approx$ 15M triples;
\item Semantic XBRL $\approx$ 100M triples;
\end{itemize}

\subsubsection{Parameter Setting.}
\label{sec:param_settings}
In order to maximise accuracy, the parameters of the algorithms have to be tweaked.
Therefore, we experimented with different parameter values and analysed the metric results.
Parameter settings were obtained by observing the algorithm's parameters in correlation with the datasets and metrics.
The rationale behind this experiment is to identify a single parameter that, when used in a metric, gives acceptable results within reasonable time.
This experiment was not performed on all datasets, since in certain cases the actual metric does not complete its computation.

The \emph{Dereferenceability} metric was implemented using reservoir sampling.
Table~\ref{tbl:deref_param} shows the time taken (in seconds) and the approximate value for different parameter settings.
The biggest time factor in this metric is the network access time, i.e. the time an HTTP request takes to respond.
The parameter settings employed for this experiment are: 
(P1) global reservoir size: 10, PLD reservoir size: 1000; 
(P2) global reservoir size: 50, PLD reservoir size: 100; 
(P3) global reservoir size: 50, PLD reservoir size: 10000; 
(P4) global reservoir size: 100, PLD reservoir size: 1000.
Whilst the approximate metrics completed the computation for all datasets, the exact computation was only ready for the LAK and LSOA datasets.
Based on the available results from the datasets, we can conclude that the optimal parameter for this metric is close to the P3 settings.
The results for the LSOA dataset are 0 due to the fact that all resources returned a \texttt{4xx/5xx} error.
This was verified manually.
\begin{table}[tb]
\centering
\resizebox{.8\textwidth}{!}{%
\begin{tabular}{l|c|c|ll|c|c|}
\cline{2-3} \cline{6-7}
                                   & \textbf{Time (s)} & \textbf{Value} &                       &               & \textbf{Time (s)} & \textbf{Value} \\ \cline{1-3} \cline{5-7} 
\multicolumn{1}{|l|}{\textbf{Actual (LAK)}} & 1423.33           & 0.045533    & \multicolumn{1}{l|}{ \hspace{1cm}} & \textbf{Actual (LSOA)} & 3189.819          & 0              \\ \cline{1-3} \cline{5-7} 
\multicolumn{1}{|l|}{P1}           & 842.082           & 0.345548    & \multicolumn{1}{l|}{} & P1            & 149.913           & 0              \\ \cline{1-3} \cline{5-7} 
\multicolumn{1}{|l|}{P2}           & 426.892           & 0.162581    & \multicolumn{1}{l|}{} & P2            & 80.748            & 0              \\ \cline{1-3} \cline{5-7} 
\multicolumn{1}{|l|}{P3}           & 618.761           & 0.057866    & \multicolumn{1}{l|}{} & P3            & 388.522           & 0              \\ \cline{1-3} \cline{5-7} 
\multicolumn{1}{|l|}{P4}           & 480.972           & 0.262936    & \multicolumn{1}{l|}{} & P4            & 316.862           & 0              \\ \cline{1-3} \cline{5-7} 
\end{tabular}
}
\caption{Dereferenceability Metric with different Parameter Settings}
\label{tbl:deref_param}
\end{table}

Another application of the Reservoir Sampling was the \emph{Existence of Links to External Data Providers}.
Table~\ref{tbl:links_params} shows the time taken (in seconds) and the estimated value for different parameter settings.
The parameter settings used to initialise the sampler were: (P1) 5,000; (P2) 10,000; (P3) 20,000; (P4) 50,000.
The results show that the approximation technique did not record any major difference up to 2M, but the technique fares better with very big datasets (c.f. Figure~\ref{fig:graph}).
One possible reason for this is that since the actual metric is not expected to fit in-memory, our implementation uses \emph{MapDB}\footnote{\url{http://www.mapdb.org}}, a pure Java database that stores memory data structures such as hash maps on disk.
It is also worth noting that all estimates gave the same value as the actual.
The reason for this is that the number of object PLDs fits in the smallest reservoir.
Therefore, since the runtime between different parameters varies a little, setting a higher or lower reservoir sampler in this case is a matter of available memory space.
If all PLDs fit in the reservoir sampler, the result is 100\% accurate.
\begin{table}[tb]
\centering
\resizebox{.8\textwidth}{!}{%
\begin{tabular}{lccllcc}
\cline{2-3} \cline{6-7}
\multicolumn{1}{l|}{\textbf{}}        & \multicolumn{1}{c|}{\textbf{Time(s)}} & \multicolumn{1}{c|}{\textbf{Value}}              & \textbf{}             & \multicolumn{1}{l|}{\textbf{}}     & \multicolumn{1}{c|}{\textbf{Time(s)}} & \multicolumn{1}{c|}{\textbf{Value}}             \\ \cline{1-3} \cline{5-7} 
\multicolumn{1}{|l|}{\textbf{Actual (LAK)}}    & \multicolumn{1}{c|}{2.271}            & \multicolumn{1}{c|}{0.000156}                 & \multicolumn{1}{l|}{} & \multicolumn{1}{l|}{\textbf{Actual (LSOA)}} & \multicolumn{1}{c|}{3.529}            & \multicolumn{1}{c|}{3.272144 $\times$ $10^6$} \\ \cline{1-3} \cline{5-7} 
\multicolumn{1}{|l|}{P1}              & \multicolumn{1}{c|}{0.578}            & \multicolumn{1}{c|}{0.000156}                 & \multicolumn{1}{l|}{} & \multicolumn{1}{l|}{P1}            & \multicolumn{1}{c|}{1.287}            & \multicolumn{1}{c|}{3.272144 $\times$ $10^6$} \\ \cline{1-3} \cline{5-7} 
\multicolumn{1}{|l|}{P2}              & \multicolumn{1}{c|}{0.481}            & \multicolumn{1}{c|}{0.000156}                 & \multicolumn{1}{l|}{} & \multicolumn{1}{l|}{P2}            & \multicolumn{1}{c|}{1.182}            & \multicolumn{1}{c|}{3.272144 $\times$ $10^6$} \\ \cline{1-3} \cline{5-7} 
\multicolumn{1}{|l|}{P3}              & \multicolumn{1}{c|}{0.466}            & \multicolumn{1}{c|}{0.000156}                 & \multicolumn{1}{l|}{} & \multicolumn{1}{l|}{P3}            & \multicolumn{1}{c|}{1.153}            & \multicolumn{1}{c|}{3.272144 $\times$ $10^6$} \\ \cline{1-3} \cline{5-7} 
\multicolumn{1}{|l|}{P4}              & \multicolumn{1}{c|}{0.444}            & \multicolumn{1}{c|}{0.000156}                 & \multicolumn{1}{l|}{} & \multicolumn{1}{l|}{P4}            & \multicolumn{1}{c|}{1.124}            & \multicolumn{1}{c|}{3.272144 $\times$ $10^6$} \\ \cline{1-3} \cline{5-7} 
                                      &                                       &                                                  &                       &                                    &                                       &                                                 \\ \cline{2-3} \cline{6-7} 
\multicolumn{1}{l|}{\textbf{}}        & \multicolumn{1}{c|}{\textbf{Time(s)}} & \multicolumn{1}{c|}{\textbf{Value}}              & \textbf{}             & \multicolumn{1}{l|}{\textbf{}}     & \multicolumn{1}{c|}{\textbf{Time(s)}} & \multicolumn{1}{c|}{\textbf{Value}}             \\ \cline{1-3} \cline{5-7} 
\multicolumn{1}{|l|}{\textbf{Actual (S'OTON)}} & \multicolumn{1}{c|}{23.872}           & \multicolumn{1}{c|}{6.166189 $\times$ $10^6$} & \multicolumn{1}{l|}{} & \multicolumn{1}{l|}{\textbf{Actual (WN)}}   & \multicolumn{1}{c|}{33.82}            & \multicolumn{1}{c|}{0}                          \\ \cline{1-3} \cline{5-7} 
\multicolumn{1}{|l|}{P1}              & \multicolumn{1}{c|}{20.693}           & \multicolumn{1}{c|}{6.166189 $\times$ $10^6$} & \multicolumn{1}{l|}{} & \multicolumn{1}{l|}{P1}            & \multicolumn{1}{c|}{7.362}            & \multicolumn{1}{c|}{0}                          \\ \cline{1-3} \cline{5-7} 
\multicolumn{1}{|l|}{P2}              & \multicolumn{1}{c|}{20.008}           & \multicolumn{1}{c|}{6.166189 $\times$ $10^6$} & \multicolumn{1}{l|}{} & \multicolumn{1}{l|}{P2}            & \multicolumn{1}{c|}{7.779}            & \multicolumn{1}{c|}{0}                          \\ \cline{1-3} \cline{5-7} 
\multicolumn{1}{|l|}{P3}              & \multicolumn{1}{c|}{20.4}             & \multicolumn{1}{c|}{6.166189 $\times$ $10^6$} & \multicolumn{1}{l|}{} & \multicolumn{1}{l|}{P3}            & \multicolumn{1}{c|}{7.557}            & \multicolumn{1}{c|}{0}                          \\ \cline{1-3} \cline{5-7} 
\multicolumn{1}{|l|}{P4}              & \multicolumn{1}{c|}{20.589}           & \multicolumn{1}{c|}{6.166189 $\times$ $10^6$} & \multicolumn{1}{l|}{} & \multicolumn{1}{l|}{P4}            & \multicolumn{1}{c|}{7.258}            & \multicolumn{1}{c|}{0}                          \\ \cline{1-3} \cline{5-7} 
\end{tabular}
}
\caption{Existence of Links to External Data Providers Metric with different Parameter Settings}
\label{tbl:links_params}
\end{table}

The \emph{Extensional Conciseness} metric was implemented using Bloom Filters.
Table~\ref{tbl:ex_cons_param} shows the time taken (in seconds) and the estimated value for different parameter settings.
We applied 4 different settings for experimentation: 
(P1) 2 filters ($k$) with a size ($M$) of 1,000; 
(P2) 5 filters with a size of 10,000; 
(P3) 10 filters with a size of 100,000; 
(P4) 15 filters with a size of 10,000,000.
This technique showed a lot of potential in the de-duplication process.
The time taken in the approximate algorithms are lower than the actual, with results being almost as accurate.
Based on the Bloom Filter trade-off, a setting between P3–P4 would exploit the potential of this technique in assessing the quality of linked datasets with regard to duplication problems.
\begin{table}[tb]
\centering
\resizebox{.8\textwidth}{!}{%
\begin{tabular}{lccllcc}
\cline{2-3} \cline{6-7}
\multicolumn{1}{l|}{}                 & \multicolumn{1}{c|}{\textbf{Time(s)}} & \multicolumn{1}{c|}{\textbf{Value}} &                       & \multicolumn{1}{l|}{}              & \multicolumn{1}{c|}{\textbf{Time(s)}} & \multicolumn{1}{c|}{\textbf{Value}} \\ \cline{1-3} \cline{5-7} 
\multicolumn{1}{|l|}{\textbf{Actual (LAK)}}    & \multicolumn{1}{c|}{81.334}           & \multicolumn{1}{c|}{0.994860}    & \multicolumn{1}{l|}{} & \multicolumn{1}{l|}{\textbf{Actual (LSOA)}} & \multicolumn{1}{c|}{375.873}          & \multicolumn{1}{c|}{1}              \\ \cline{1-3} \cline{5-7} 
\multicolumn{1}{|l|}{P1}              & \multicolumn{1}{c|}{1.348}            & \multicolumn{1}{c|}{0.621315}    & \multicolumn{1}{l|}{} & \multicolumn{1}{l|}{P1}            & \multicolumn{1}{c|}{1.043}            & \multicolumn{1}{c|}{0.617729}    \\ \cline{1-3} \cline{5-7} 
\multicolumn{1}{|l|}{P2}              & \multicolumn{1}{c|}{1.377}            & \multicolumn{1}{c|}{0.962249}     & \multicolumn{1}{l|}{} & \multicolumn{1}{l|}{P2}            & \multicolumn{1}{c|}{1.328}            & \multicolumn{1}{c|}{0.966795}    \\ \cline{1-3} \cline{5-7} 
\multicolumn{1}{|l|}{P3}              & \multicolumn{1}{c|}{1.67}             & \multicolumn{1}{c|}{0.993946}    & \multicolumn{1}{l|}{} & \multicolumn{1}{l|}{P3}            & \multicolumn{1}{c|}{1.807}            & \multicolumn{1}{c|}{0.999240}    \\ \cline{1-3} \cline{5-7} 
\multicolumn{1}{|l|}{P4}              & \multicolumn{1}{c|}{2.212}            & \multicolumn{1}{c|}{0.994593}    & \multicolumn{1}{l|}{} & \multicolumn{1}{l|}{P4}            & \multicolumn{1}{c|}{2.98}             & \multicolumn{1}{c|}{1}              \\ \cline{1-3} \cline{5-7} 
                                      &                                       &                                     &                       &                                    &                                       &                                     \\ \cline{2-3} \cline{6-7} 
\multicolumn{1}{l|}{}                 & \multicolumn{1}{c|}{\textbf{Time(s)}} & \multicolumn{1}{c|}{\textbf{Value}} &                       & \multicolumn{1}{l|}{}              & \multicolumn{1}{c|}{\textbf{Time(s)}} & \multicolumn{1}{c|}{\textbf{Value}} \\ \cline{1-3} \cline{5-7} 
\multicolumn{1}{|l|}{\textbf{Actual (S'OTON)}} & \multicolumn{1}{c|}{7366.225}         & \multicolumn{1}{c|}{0.737523}     & \multicolumn{1}{l|}{} & \multicolumn{1}{l|}{\textbf{Actual (WN)}}   & \multicolumn{1}{c|}{96511.334}       & \multicolumn{1}{c|}{0.948}     \\ \cline{1-3} \cline{5-7} 
\multicolumn{1}{|l|}{P1}              & \multicolumn{1}{c|}{24.304}           & \multicolumn{1}{c|}{0.512887}    & \multicolumn{1}{l|}{} & \multicolumn{1}{l|}{P1}            & \multicolumn{1}{c|}{7.407}            & \multicolumn{1}{c|}{0.570991}    \\ \cline{1-3} \cline{5-7} 
\multicolumn{1}{|l|}{P2}              & \multicolumn{1}{c|}{20.217}           & \multicolumn{1}{c|}{0.782946}    & \multicolumn{1}{l|}{} & \multicolumn{1}{l|}{P2}            & \multicolumn{1}{c|}{11.502}           & \multicolumn{1}{c|}{0.885790}    \\ \cline{1-3} \cline{5-7} 
\multicolumn{1}{|l|}{P3}              & \multicolumn{1}{c|}{17.512}           & \multicolumn{1}{c|}{0.783529}    & \multicolumn{1}{l|}{} & \multicolumn{1}{l|}{P3}            & \multicolumn{1}{c|}{17.653}           & \multicolumn{1}{c|}{0.900407}    \\ \cline{1-3} \cline{5-7} 
\multicolumn{1}{|l|}{P4}              & \multicolumn{1}{c|}{20.275}           & \multicolumn{1}{c|}{0.660193}    & \multicolumn{1}{l|}{} & \multicolumn{1}{l|}{P4}            & \multicolumn{1}{c|}{35.381}           & \multicolumn{1}{c|}{0.844733}    \\ \cline{1-3} \cline{5-7} 
\end{tabular}
}
\caption{Extensional Conciseness Metric with different Parameter Settings}
\label{tbl:ex_cons_param}
\end{table}

For the \emph{clustering coefficient} metric we multiplied the base mixing time of $\log^2n$ with 0.1, 0.5, 0.7 and 1.0 respectively to test with fast mixing time.
Table~\ref{tbl:clus_coef_param} shows the time taken (in seconds) and the estimated value for different parameter settings.
The results show that for the assessed datasets the $\log^2n$ mixing time is not ideal.
This is due to the fact that the smallest multiplier setting, i.e. 0.1, proved to be the closest to the actual result in all cases.
Determining a more accurate average mixing time, and hence a more accurate estimate (cf. Section~\ref{sec:preliminaries}), requires the evaluation (such as in \cite{Mohaisen2010}) of all datasets in the LOD Cloud. 
\begin{table}[tb]
\centering
\resizebox{.8\textwidth}{!}{%
\begin{tabular}{lccllcc}
\cline{2-3} \cline{6-7}
\multicolumn{1}{l|}{\textbf{}}        & \multicolumn{1}{c|}{\textbf{Time(s)}} & \multicolumn{1}{c|}{\textbf{Value}} & \textbf{}             & \multicolumn{1}{l|}{\textbf{}}     & \multicolumn{1}{c|}{\textbf{Time(s)}} & \multicolumn{1}{c|}{\textbf{Value}} \\ \cline{1-3} \cline{5-7} 
\multicolumn{1}{|l|}{\textbf{Actual (LAK)}}    & \multicolumn{1}{c|}{42.729}           & \multicolumn{1}{c|}{0.961040}    & \multicolumn{1}{l|}{} & \multicolumn{1}{l|}{\textbf{Actual (LSOA)}} & \multicolumn{1}{c|}{62.618}           & \multicolumn{1}{c|}{1}              \\ \cline{1-3} \cline{5-7} 
\multicolumn{1}{|l|}{Mixing time 0.1}             & \multicolumn{1}{c|}{4.595}            & \multicolumn{1}{c|}{0.978220}    & \multicolumn{1}{l|}{} & \multicolumn{1}{l|}{Mixing time 0.1}           & \multicolumn{1}{c|}{7.657}            & \multicolumn{1}{c|}{0.999995}    \\ \cline{1-3} \cline{5-7} 
\multicolumn{1}{|l|}{Mixing time 0.5}             & \multicolumn{1}{c|}{4.595}            & \multicolumn{1}{c|}{0.997945}    & \multicolumn{1}{l|}{} & \multicolumn{1}{l|}{Mixing time 0.5}           & \multicolumn{1}{c|}{6.829}            & \multicolumn{1}{c|}{0.999999}    \\ \cline{1-3} \cline{5-7} 
\multicolumn{1}{|l|}{Mixing time 0.7}             & \multicolumn{1}{c|}{4.766}            & \multicolumn{1}{c|}{0.998665}    & \multicolumn{1}{l|}{} & \multicolumn{1}{l|}{Mixing time 0.7}           & \multicolumn{1}{c|}{6.561}            & \multicolumn{1}{c|}{0.999503}    \\ \cline{1-3} \cline{5-7} 
\multicolumn{1}{|l|}{Mixing time 1.0}             & \multicolumn{1}{c|}{4.832}            & \multicolumn{1}{c|}{0.998974}    & \multicolumn{1}{l|}{} & \multicolumn{1}{l|}{Mixing time 1.0}           & \multicolumn{1}{c|}{6.528}            & \multicolumn{1}{c|}{0.999999}    \\ \cline{1-3} \cline{5-7} 
                                      &                                       &                                     &                       &                                    &                                       &                                     \\ \cline{2-3} \cline{6-7} 
\multicolumn{1}{l|}{}                 & \multicolumn{1}{c|}{\textbf{Time(s)}} & \multicolumn{1}{c|}{\textbf{Value}} &                       & \multicolumn{1}{l|}{}              & \multicolumn{1}{c|}{\textbf{Time(s)}} & \multicolumn{1}{c|}{\textbf{Value}} \\ \cline{1-3} \cline{5-7} 
\multicolumn{1}{|l|}{\textbf{Actual (S'OTON)}} & \multicolumn{1}{c|}{408.358}          & \multicolumn{1}{c|}{0.933590}    & \multicolumn{1}{l|}{} & \multicolumn{1}{l|}{\textbf{Actual (WN)}}   & \multicolumn{1}{c|}{9012.454}         & \multicolumn{1}{c|}{0.759257}    \\ \cline{1-3} \cline{5-7} 
\multicolumn{1}{|l|}{Mixing time 0.1}             & \multicolumn{1}{c|}{46.373}           & \multicolumn{1}{c|}{0.993067}    & \multicolumn{1}{l|}{} & \multicolumn{1}{l|}{Mixing time 0.1}           & \multicolumn{1}{c|}{243.009}          & \multicolumn{1}{c|}{0.810405}    \\ \cline{1-3} \cline{5-7} 
\multicolumn{1}{|l|}{Mixing time 0.5}             & \multicolumn{1}{c|}{46.362}           & \multicolumn{1}{c|}{0.997634}    & \multicolumn{1}{l|}{} & \multicolumn{1}{l|}{Mixing time 0.5}           & \multicolumn{1}{c|}{248.925}          & \multicolumn{1}{c|}{0.999919}    \\ \cline{1-3} \cline{5-7} 
\multicolumn{1}{|l|}{Mixing time 0.7}             & \multicolumn{1}{c|}{46.238}           & \multicolumn{1}{c|}{0.997939}    & \multicolumn{1}{l|}{} & \multicolumn{1}{l|}{Mixing time 0.7}           & \multicolumn{1}{c|}{251.396}          & \multicolumn{1}{c|}{0.999917}    \\ \cline{1-3} \cline{5-7} 
\multicolumn{1}{|l|}{Mixing time 1.0}             & \multicolumn{1}{c|}{46.225}           & \multicolumn{1}{c|}{0.998312}    & \multicolumn{1}{l|}{} & \multicolumn{1}{l|}{Mixing time 1.0}           & \multicolumn{1}{c|}{252.522}          & \multicolumn{1}{c|}{0.999967}    \\ \cline{1-3} \cline{5-7} 
\end{tabular}
}
\caption{Clustering Coefficient Metric with different Parameter Settings}
\label{tbl:clus_coef_param}
\end{table}

\subsubsection{Evaluation Discussion.}
\begin{figure}[tb]
  \centering
   \includegraphics[width=\textwidth]{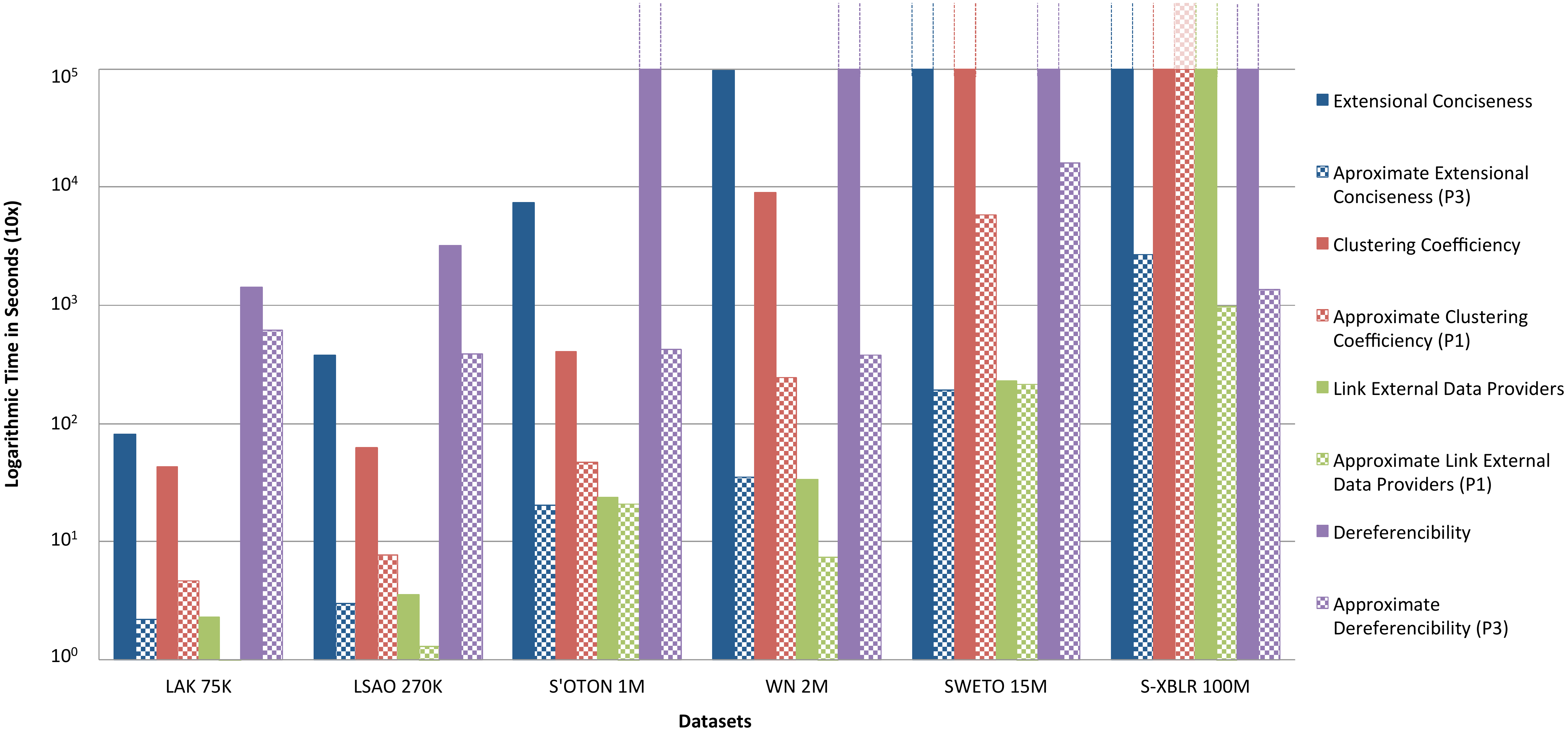} 
   \caption{Runtime of Metrics vs. Datasets}
   \label{fig:graph}
\end{figure}

Our experiments gave promising results towards the use and acceptance of probabilistic approximation for estimating the quality of linked open datasets.
Figure~\ref{fig:graph} shows the time taken in all implemented metrics (actual and approximated) against the evaluated datasets.
The graph clearly shows that all approximate metrics have a lower runtime than their equivalent actual metric.
Whilst the approximate metrics for \emph{link external data providers}, \emph{dereferenceability}, and \emph{extensional conciseness} computed all metrics, the approximate \emph{clustering coefficient} and the actual \emph{link external data providers} managed to compute 5 datasets within a reasonable time. 
The actual \emph{dereferenceability} metric managed only to compute two datasets, while the other two actual metrics computed up to the WordNet dataset.
Table~\ref{tbl:values} shows the metric (actual and estimated) values for the datasets.
\begin{table}[tb]
\centering
\resizebox{\textwidth}{!}{%
\begin{tabular}{l|l|l|l|l|l|l|}
\cline{2-7}
                                                                        & \textbf{LAK 75K}              & \textbf{LSOA 270K}              & \textbf{S'OTON 1M}           & \textbf{WN 2M} & \textbf{SWETO 15M} & \textbf{S-XBLR 100M}          \\ \hline
\multicolumn{1}{|l|}{\textbf{Extensional Conciseness}}                  & 0.9948                   & 1                               & 0.7375                   &    0.948            & 0,000370             & N/A                           \\ \hline
\multicolumn{1}{|l|}{\textbf{Approx. Extensional Conciseness}}       & 0.9945                   & 1                               & 0.6601                  & 0.8447    & 0.9998        & 0.1097                    \\ \hline
\multicolumn{1}{|l|}{\textbf{Clustering Coefficiency}}                  & 0.9610                   & 1                               & 0.9335                  & 0.7592    & N/A                & N/A                            \\ \hline
\multicolumn{1}{|l|}{\textbf{Approx. Clustering Coefficiency}}      & 0.9782                   & 0.9999                     & 0.9930                  & 0.8104    & 1                  & 0                             \\ \hline
\multicolumn{1}{|l|}{\textbf{Link External Data Providers}}             & 0.01569$\times$$10^{-6}$ & 3.2721$\times$$10^{-6}$  & 6.1661$\times$$10^{-6}$ & 0              & N/A                & N/A                           \\ \hline
\multicolumn{1}{|l|}{\textbf{Approx. Link External Data Prov.}} & 0.01569$\times$$10^{-6}$ & 3.2721$\times$$10^{-6}$ & 6.1661$\times$$10^{-6}$ & 0              & 0.000370        & 4,9557$\times$$10^{-8}$                     \\ \hline
\multicolumn{1}{|l|}{\textbf{Dereferencibility}}                        & 0.0455                   & 0                               & N/A                          & N/A            & N/A                & N/A                          \\ \hline
\multicolumn{1}{|l|}{\textbf{Approx. Dereferencibility}}            & 0.0578                   & 0                               & 0.4122                  & 0.9681    & 0                  & 0,0955$\times$$10^{-8}$ \\ \hline
\end{tabular}
}
\caption{Metric value (Actual and Approximate) per dataset}
\label{tbl:values}
\end{table}
The approximate results are in most cases very close to the actual results.
However, approximate measures are calculated in an acceptable time unlike their actual counterparts.
As part of a larger effort to implement scalable LOD quality assessment metrics, we assessed the metrics identified in~\cite{Zaveri2012:LODQ} and assigned to them possible approximation techniques discussed in this article (cf. Table~\ref{tbl:big_data_impl}).
\begin{table}[h]
\centering
\resizebox{\textwidth}{!}{%
\begin{tabular}{|l|c|c|l|}
\hline
\textbf{Metric}                              & \textbf{Approximation Technique} \\ \hline
Dereferenceability of the URI                        & Reservoir Sampling            \\ \hline
Dereferenced Forward-Links                                & Reservoir Sampling                           \\ \hline
Detection of Good Quality Interlinks            & Random Walk                   \\ \hline
Dereferenced Back-Links                             & Reservoir Sampling            \\ \hline
Usage of Slash-URIs                            & Reservoir Sampling            \\ \hline
Syntactically Accurate Values                          & Reservoir Sampling            \\ \hline
No Misuse of Properties                             & Reservoir Sampling            \\ \hline
No Use of Entities as Members of Disjoint Classes           & Reservoir Sampling            \\ \hline
High Extensional Conciseness                          & Bloom Filters                 \\ \hline
High Intensional Conciseness                            & Bloom Filters                 \\ \hline
Duplicate Instance                                & Bloom Filters                 \\ \hline
Relevant Terms Within Meta-Information Attributes       & Page Rank              \\ \hline
Coverage                                               & Reservoir Sampling            \\ \hline
\end{tabular}
}
\caption{Possible Metric  Approximation Implementation}
\label{tbl:big_data_impl}
\end{table}

Overall, given that the results were obtained on yet small datasets (the chosen ones might not be considered to be big enough) due to limited infrastructure, this paper contributes towards invaluable results that can be the basis for further studies.
These results show that with probabilistic approximation techniques:
\begin{enumerate}
	\item Runtime decreases considerably -- for larger datasets easily by more than an order of magnitude;
	\item Loss of precision is acceptable in most cases with less than 10\% deviation from actual values;
	\item Large linked datasets can be assessed for quality even within very limited computational capabilities, such as a personal notebook.
\end{enumerate} 



\section{Conclusion}
\label{sec:conclusion}


In this article, we have demonstrated how the three approximate techniques reservoir sampling, Bloom Filters and clustering coefficient estimation can be successfully applied for Linked Data quality assessment.
Our comprehensive experiments have shown that we can reduce runtime in most cases by more than an order of magnitude, while keeping the precision of results reasonable for most practical applications.
All in all, we have demonstrated that using these approximation techniques enables data publishers to assess their datasets in a convenient and efficient manner without the need of having a large infrastructure for computing quality metrics.
Therefore, data publishers are encouraged to assess their data before publishing it to the Web, thus ensuring that data consumers receive quality data at their end.

In terms of Linked Data quality assessment we aim to extend our work both in terms of used big data techniques and metric coverage.
Regarding probabilistic approximation techniques, we aim to assess other probabilistic data structures such as quotient filters or random trees.
A further interesting avenue of research is to investigate how such techniques can be easily employed for domain specific data quality metrics.

\bibliographystyle{splncs03}
\bibliography{paper,../../bib/eis,../../bib/all,../../bib/external/kwarc} 

\end{document}